\documentclass[10pt,conference,a4paper]{IEEEtran}
\IEEEoverridecommandlockouts

\usepackage{cite}
\usepackage{amsmath,amssymb,amsfonts}
\usepackage{booktabs}
\usepackage{graphicx}
\usepackage{textcomp}
\usepackage{xcolor}
\usepackage{url}
\usepackage{hyperref}
\usepackage{tcolorbox}
\usepackage{tikz}
\usepackage{ifthen}
\usepackage{xspace}
\definecolor{withcontext}{HTML}{4C78A8}
\definecolor{withoutcontext}{HTML}{F58518}

\hypersetup{hidelinks}

\author{\IEEEauthorblockN{Kaya Kitada, Yoshiki Higo, Raula Gaikovina Kula}
\IEEEauthorblockA{The University of Osaka, Japan\\
\{ka-kitad, higo, raula-k\}@ist.osaka-u.ac.jp}}

\def\BibTeX{{\rm B\kern-.05em{\sc i\kern-.025em b}\kern-.08em
    T\kern-.1667em\lower.7ex\hbox{E}\kern-.125emX}}

\begin{document}

\title{
Agentic Dependency Transformations:\\
Lessons from an Exploratory Case Study of \texttt{Flask}
}
\maketitle

\begin{abstract}
Modern software relies heavily on third-party libraries, requiring maintainers to balance the benefits of reuse against dependency-related maintenance and risk. We define \emph{dependency transformation} as an emerging approach in which an AI coding agent alters how a repository handles external dependencies. Local reimplementation is one transformation strategy that recreates required functionality as local code; dependency transformation is broader and may include other strategies.

In this emerging-results paper, we report an exploratory case study of Flask to explore dependency transformation strategies.
We design experiments that show how available context shapes an agent's dependency discovery and iterative transformation of dependencies.
The case yields five preliminary lessons.
These lessons and the five-part research agenda position dependency transformation as an emerging research direction and motivate explicit transformation contracts, provenance constraints, preservation requirements, and independent validation criteria.
\end{abstract}


\section{Introduction}
Modern software is assembled from ecosystems of external libraries and frameworks such as npm, PyPI, and Maven. Reuse accelerates delivery, but it also creates update work, version conflicts, vulnerabilities, and supply-chain exposure~\cite{Analyzing-Maintenance,Kula2017,Surviving-software-dependencies}. A 2024 Synopsys report found open-source software in 96\% of the audited commercial codebases and estimated that 77\% of their code was open source.\footnote{\href{https://www.blackduck.com/resources/analyst-reports/open-source-security-risk-analysis.html}{Synopsys OSSRA 2024}} Widely propagated vulnerabilities such as Log4Shell illustrate how risk in a single library can affect many downstream systems.\footnote{\href{https://github.blog/open-source/inside-the-breach-that-broke-the-internet-the-untold-story-of-log4shell/}{GitHub, ``The Untold Story of Log4Shell''}}.
Research has also shown dependency management impacts security \cite{pashchenko2020qualitative,Research-Directions}.

Existing tools support dependency identification, vulnerability scanning, and version maintenance~\cite{Automated-dependency,Generating-precise,An-Overview-and-Catalogue-of-Dependency}. For example, Dependabot automates update proposals and vulnerability notifications. Developers must nevertheless determine where and why a dependency is used, whether updating or removing it is safe, and whether a proposed change preserves repository behavior.

The idea of using Large Language Models (LLMs) for dependency management is now in early exploration. 
For instance, Almeida et. al,\cite{almeida2024automatic} and Tawosi et al. ~\cite{tawosi2025llm} both explore how dependency recommendations by an LLM. 
Related research addresses automation for dependency identification, vulnerability analysis, resolution, and automated updates~\cite{Automated-dependency,Generating-precise,An-Overview-and-Catalogue-of-Dependency,Research-Directions,jafari2023dependency,he2023automating}. 
Apart from updates, dependency strategies may involve dependency downgrades\cite{cogo2019empirical} and removals \cite{jaisri2025preliminary} for self-containment.

An AI coding agent may complement these dependency bots by searching a repository, inspecting heterogeneous artifacts, editing multiple files, and running builds and tests~\cite{Agentic-Software-Engineering}. We subsequently refer to the AI coding agent as the \emph{agent}. These capabilities motivate \emph{dependency transformation}, which which we define as how an agent alters how a repository handles external dependencies.

We define a dependency transformation as a repository-level change in which an agent takes one third-party package that the repository relies on and changes how the functionality obtained from that package is supplied, so that the repository no longer needs the package.
A transformation therefore spans three kinds of edits: (i) the code that imports or calls the package, (ii) the files that declare it, such as manifests, lockfiles, and CI/CD configuration, and (iii) the build and tests used to validate the dependency.
For example, if a repository imports a package only to load a configuration file, an agent may rewrite that loader locally, delete the package from \texttt{pyproject.toml}, and re-run the tests.
This differs from a dependency update, which keeps the package and changes its version, and from a library migration, which is when there is a replacement of one third-party package with another.
Local reimplementation, as in the example, is one transformation strategy; dependency transformation is broader and also covers strategies such as using the standard library, copying the package source into the repository, or removing the feature that needs the package.
We call a third-party package that the agent reports as a dependency of the repository, and that is therefore a potential target of a transformation, a \emph{dependency-transformation candidate}.

In this paper, we study dependency transformation in a real-world setting using the agent rather than a standalone LLM.
We explore a case study of Flask using OpenAI ChatGPT~5.4\footnote{\href{https://openai.com/index/introducing-gpt-5-4/}{OpenAI GPT-5.4}} and the pi coding-agent harness.\footnote{\href{https://github.com/earendil-works/pi}{pi coding agent}}.
Rather than establish effectiveness, our study uses one repository-level case to derive two research questions:

\begin{itemize}
    \item \textbf{RQ1:} \textit{How does available context shape the discovery of dependency-transformation candidates?}\\
    The motivation is to compare the agent's dependency discovery with and without context (only source code).
    
    \item \textbf{RQ2:} \textit{What transformation strategies arise when an agent iteratively processes transformation candidates at the repository level?}\\
    In addition to local reimplementation, we explore other strategies suggested and implemented by the agent.
\end{itemize}

\section{Study Design}
\label{sec:method}

Flask was selected because it is well studied in software engineering and represents a typical open source project that contains all artifacts such as runtime code, tests, documentation, examples, packaging metadata, and CI/CD configuration. 
The experiments used the same fixed Flask commit, \texttt{2ac89889f4cc} (236 files; 732 contributors, 5,537 commits, and 2,791 pull requests).

We ran the agent with OpenAI ChatGPT~5.4 once per condition through the open-source pi harness.
A custom extension recorded observable actions, including opened files, search commands, tool invocations, dialogue turns, model requests, token counts, and generated modifications. These are harness-boundary observations, not the model's private reasoning.
We executed our experiments on May 9--10, 2026. Prompts, withheld files, per-step commands, and guardrail files are documented in the replication package.

\section{Empirical Study}

\subsection{Answering RQ1}
To answer RQ1, we designed two experiments to compare dependency discovery with and without context.
We controlled the context by running separate sessions with or without manifest files.

\paragraph{Controlling Context}
We define context as the manifest files\footnote{These include files such as \texttt{pyproject.toml}, \texttt{uv.lock}, and \texttt{requirements.txt}.}. These files are not complete ground truth: they can contain optional, transitive, test, documentation, tooling, workflow, or example-specific packages.
We assumed that without context, the agent relied upon imports, tests, documentation, examples, and workflows within the code and other non-manifest files.

\paragraph{Agent Prompt}
Using an agent mode, we provide a broad instruction to \emph{``List all third-party dependencies.''}. 
The agent then autonomously chose its own search path and could inspect source, tests, documentation, examples, and workflows.

\paragraph{Analysis of Results}
We collected every third-party package the agent reported, excluding standard-library modules.

\paragraph{Classification of Candidates}
A candidate count alone does not show whether the two sessions differ in how well they discover dependencies or only in which packages they treat as dependencies.
To separate the two, the first author manually classified each candidate reported with context by where it originates in the repository.
Each candidate was assigned to exactly one of five categories, checked in this order:
(1) \emph{root declaration}, declared by the root \texttt{pyproject.toml} as a runtime, optional, dependency-group, or build-system requirement;
(2) \emph{lockfile}, appearing only as a resolved transitive entry in \texttt{uv.lock};
(3) \emph{example}, required only by a project under the examples directory;
(4) \emph{documentation}, only mentioned in documentation, tutorials, or deployment guides; and
(5) \emph{build tooling}, only referenced as a tool used to build the project.
For each session, we also recorded elapsed time, turns, tool calls, and tokens.

\begin{table}[t]
\centering
\scriptsize
\setlength{\tabcolsep}{2.5pt}
\begin{minipage}[t]{.47\linewidth}
\centering
\caption{Agent executions with and without context}
\label{tab:rq1-log-summary}
\begin{tabular}{lrr}
\toprule
Metric & With & Without \\
\midrule
Execution time (s) & 247.4 & 266.0 \\
Dialogue turns & 10 & 24 \\
Tool executions & 45 & 97 \\
Input tokens & 153,323 & 70,178 \\
Output tokens & 17,614 & 16,146 \\\midrule
\texttt{read} & 31 & 29 \\
\texttt{grep} & 11 & 45 \\
\texttt{ls} & 1 & 12 \\
\texttt{find} & 1 & 10 \\
\texttt{write} & 1 & 1 \\
\bottomrule
\end{tabular}
\end{minipage}\hfill
\begin{minipage}[t]{.51\linewidth}
\centering
\caption{Candidates reported with context by primary source ($n=114$)}
\label{tab:rq1-classification}
\begin{tabular}{lr}
\toprule
Category & Cand. \\
\midrule
Root declaration & 27 \\
Lockfile (transitive only) & 54 \\
Example & 14 \\
Documentation & 18 \\
Build tooling & 1 \\
\midrule
Total & 114 \\
\bottomrule
\end{tabular}
\end{minipage}
\end{table}

\begin{figure*}[t]
\centering
\includegraphics[width=.82\linewidth]{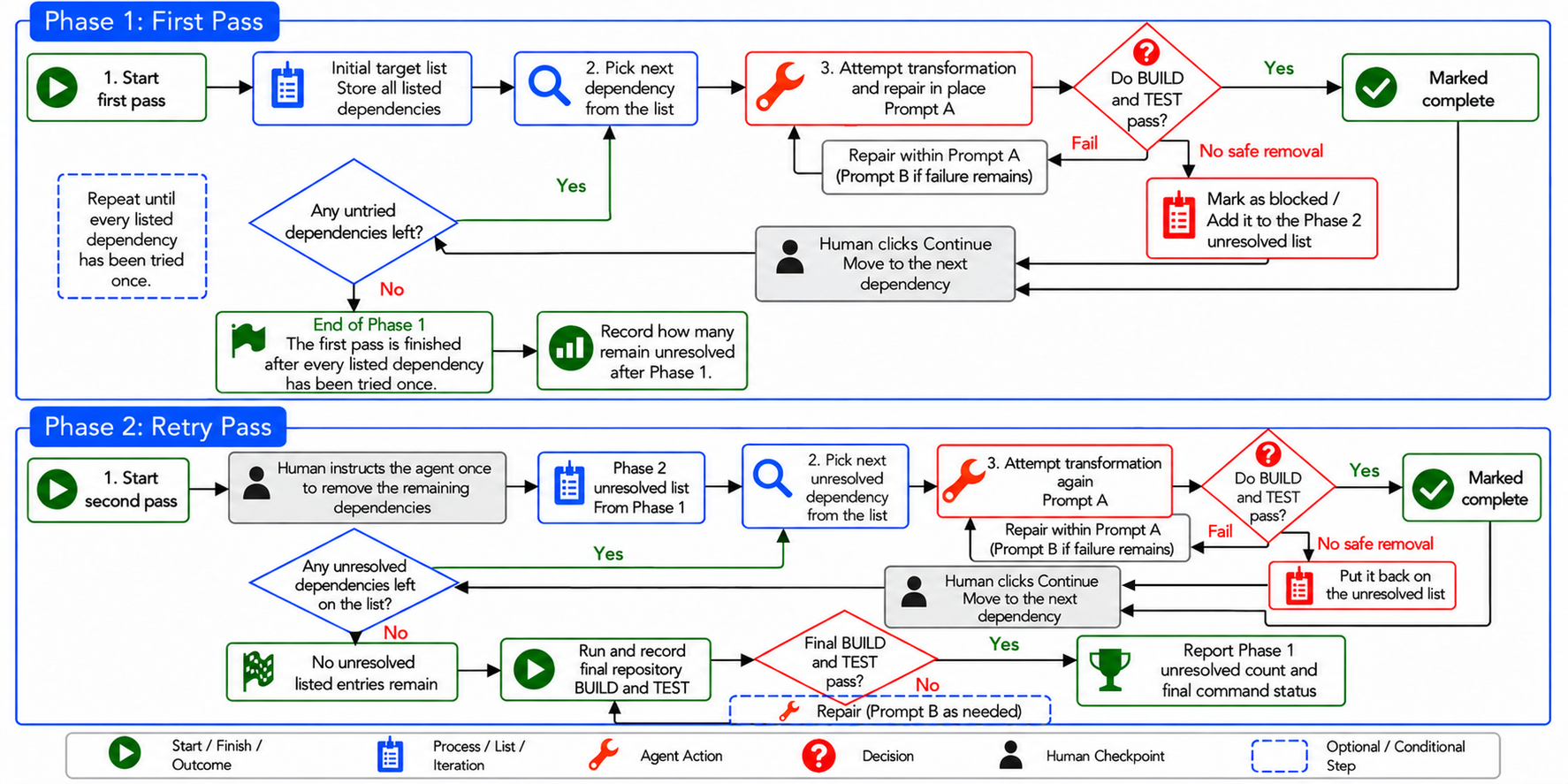}
\caption{Dependency transformation workflow for RQ2}
\label{fig:workflow}
\end{figure*}

\subsection{RQ1 Results}
Table~\ref{tab:rq1-log-summary} shows the cost of each session.
Execution times are similar, but with context the agent reads the manifests and uses more input tokens, whereas without context it searches the repository more often with \texttt{grep}, \texttt{ls}, and \texttt{find}.

Of the 115 distinct candidates, 27 were reported in both sessions, 87 only with context, and 1 only without context.
Table~\ref{tab:rq1-classification} classifies the 114 candidates reported with context: only 27 (24\%) are root declarations, whereas 54 (47\%) are lockfile-only entries.
The larger count with context is therefore mostly explained by a wider reported scope, driven by the lockfile, rather than by more of the directly declared dependencies being found: the session without context still recovered 19 of the 27 root declarations from imports, tests, and documentation.

\begin{tcolorbox}[title=Answer to RQ1]
Context changes which packages the agent reports as dependencies: 114 candidates with context, mostly lockfile-only entries, against 28 without.
This is a difference in reported scope, not evidence of better discovery, since each condition was run once and there is no ground truth to measure precision or recall.
\end{tcolorbox}

\subsection{Answering RQ2}

To answer RQ2, we designed a dependency transformation workflow that applies transformations to transformation candidates.

\paragraph{Dependency Transformation Workflow}
Figure \ref{fig:workflow} provides an overview of the dependency transformation workflow.
The workflow is driven by an orchestration prompt that makes the agent track the remaining transformation candidates and process them one at a time, using two prompts:
\begin{itemize}
    \item \textit{Prompt A} instructs the agent to perform a transformation.
    The prompt instructs the agent to remove one transformation candidate without adding a new third-party dependency, provide needed functionality locally when applicable, and run the build and tests.
    If the build or tests fail, the agent repairs the change and validates again within the same prompt.
    \item \textit{Prompt B} is a repair prompt that takes the failure log of a failing build or test and applies a minimal fix. It is used only if a failure remains after Prompt A or appears in the final validation.
\end{itemize}
Each experiment ran as one continuous agent session, so the agent kept its context across candidates; the two experiments ran in separate sessions.
We divide the workflow into two phases:
\begin{itemize}
    \item \textbf{First Pass}---The agent applies Prompt A once to each transformation candidate. A candidate that the agent cannot remove safely is recorded as blocked and remains in the list of remaining candidates.
    \item \textbf{Retry Pass}---When the agent ended the First Pass with blocked candidates, the first author instructed it once to remove the remaining candidates. The agent then applied Prompt A again to each blocked candidate.
\end{itemize}

\paragraph{Termination}
The workflow sets neither a maximum number of retries nor a timeout.
An attempt on one candidate ends when the agent judges the transformation complete or judges that no further useful progress can be made, in which case the candidate is recorded as blocked.
The whole process ends as successful only when no candidate remains, no new third-party dependency has been introduced, and a final build and test run passes; otherwise the agent must record the run as incomplete together with the reason.

\begin{table}[t]
\centering
\scriptsize
\setlength{\tabcolsep}{2.5pt}
\begin{minipage}[t]{.51\linewidth}
\centering
\caption{Completions by phase and type}
\label{tab:completion}
\begin{tabular}{lrrr}
\toprule
 & With & W/o & Total \\
\midrule
Candidates processed & 43 & 21 & 64 \\
Completed, First Pass & 20 & 17 & 37 \\
Completed, Retry Pass & 23 & 4 & 27 \\
\midrule
Direct completion & 31 & 21 & 52 \\
Cascading completion & 12 & 0 & 12 \\
\bottomrule
\end{tabular}
\end{minipage}\hfill
\begin{minipage}[t]{.47\linewidth}
\centering
\caption{Audit of test artifacts}
\label{tab:test-audit}
\begin{tabular}{lrr}
\toprule
Audit & With & W/o \\
\midrule
With test edits & 38/43 & 19/21 \\
New test files & 16 & 15 \\
Existing files modified & 3 & 2 \\
Deleted test files & 0 & 0 \\
\bottomrule
\end{tabular}
\end{minipage}
\end{table}

\begin{table}[t]
\centering
\caption{Strategies used for direct completions ($n=52$).}
\label{tab:outcomes}
\scriptsize
\begin{tabular}{p{.52\linewidth}rrr}
\toprule
Transformation Strategy & With & Without & Total \\
\midrule
Local reimplementation & 4 & 5 & 9 \\
Standard-library implementation & 3 & 1 & 4 \\
Copy-and-own & 9 & 4 & 13 \\
Dependency removal & 7 & 10 & 17 \\
Mixed/configuration replacement & 8 & 1 & 9 \\
\midrule
Total & 31 & 21 & 52 \\
\bottomrule
\end{tabular}
\end{table}

\paragraph{Candidates and Processing Order}
We ran two transformation experiments, one with context and one without.
The candidates were not sampled from the RQ1 lists: each RQ2 session began with its own discovery step, now with the goal of removing dependencies, and the agent processed every candidate it listed, 43 with context and 21 without.
All 43 are among the 114 candidates of RQ1: the 27 root declarations, the 14 example candidates, and 2 lockfile-only packages that the documentation code imports.
The agent left out the other 52 lockfile-only candidates and the 19 that are only mentioned in documentation or build tooling.
Of the 21 candidates without context, 20 are among the 28 of RQ1.
The agent processed candidates one after another on the same working copy, in an order it chose; with context, it removed loosely coupled dependencies before core runtime ones.
Changes therefore accumulate, and the outcome for one candidate can depend on earlier transformations.
We did not vary the order or reset the repository between candidates.

\paragraph{Completion Criterion and Human Checkpoint}
A candidate reaches \emph{transformation completion} when it is removed from the repository and the build and tests pass.
The tests are not frozen: Prompt A lets the agent update or add tests when needed.
The first author monitored each agent execution and checked that each transformation was carried out as instructed (Figure~\ref{fig:workflow}).

\subsection{RQ2 Results}

\paragraph{Completion}
Table~\ref{tab:completion} summarizes completion. All 64 candidates reached transformation completion: 37 in the First Pass and the remaining 27 in the Retry Pass.
Twelve completions, all with context, were \emph{cascading}: the candidate was resolved as a side effect of transforming another one (all 12 were dependencies of the Celery example and were resolved when Celery was removed).
We focus on the other 52 \emph{direct} completions (31 with context, 21 without).
The experiment with context took 3.97 hours, 54 harness runs, and 1,068 agent turns; the one without took 1.40 hours, 23 runs, and 403 turns.

\paragraph{A Taxonomy of Dependency Transformation Strategies}
Table~\ref{tab:outcomes} reports the transformation strategy for the 52 direct completions.
We define the strategies as follows:

\begin{enumerate}
    \item \textbf{Local reimplementation}
    Reimplemented the required functionality directly within the repository.
    \item \textbf{Standard-library implementation}
    Replaced the external dependency with equivalent functionality from the standard library.
    \item \textbf{Copy-and-own}
    Copied the dependency’s source code into the repository and used it locally.
    \item \textbf{Dependency removal}
    Removed or simplified the feature or development tool that required the dependency.
    \item \textbf{Mixed/configuration replacement}
    \begin{itemize}
        \item 
        Removed from optional requirements, configuration, or metadata.
        \item 
        Resolved through a combination of code changes, configuration edits, and other strategies.
    \end{itemize}
\end{enumerate}

As shown in the table, dependency removal was the most common strategy, followed by copy-and-own and local reimplementation.

\paragraph{Retries and Repairs}
One retry instruction per experiment was sufficient, and no candidate was attempted more than twice.
The 27 blocked candidates were not build or test failures: the agent stopped because it found no safe way to remove the candidate, and it left the build and tests passing.
Build or test failures were repaired within Prompt A (11 of the 53 attempts with context, at most two repair cycles each; none without context).
One test failed in the final validation without context, and the agent fixed it directly.
Prompt B was therefore not invoked in either experiment.

\paragraph{Build and Test Artifacts}
Table~\ref{tab:test-audit} audits the test files. The agent edited tests in 57 of the 64 transformations, created 31 new test files, and modified 5 existing ones; it deleted none.
At least one transformation changed an original behavioral expectation.
Completion as defined above is therefore a weak criterion: it shows that the candidate is gone and that the current, partly agent-written, tests pass, not that the original behavior is preserved.

\begin{tcolorbox}[title=Answer to RQ2]
We identify five dependency transformation strategies for Flask; only 13 of the 52 direct completions reimplemented functionality locally or with the standard library.
All candidates reached completion, but the agent edited tests in most transformations, so completion does not establish behavioral preservation.
\end{tcolorbox}

\section{Limitations and Lessons Learnt}
\paragraph{Limitations and Threats to Validity}
\label{sec:limitations}
First, this is one case study, specific to Flask's structure, tests, and dependency practices.
Second, each condition was run once with a stochastic agent, so differences between conditions cannot be separated from run-to-run variation.
Third, we have no ground truth of Flask's dependencies and no second annotator, so we report neither precision nor recall, and the classification labels are descriptive.
Fourth, the RQ2 workflow is cumulative: candidates were processed in an agent-chosen order on one working copy, so order and cascading effects are confounded with the difficulty of each candidate. Isolated, per-candidate runs may give different results.
Fifth, completion relies on tests that the agent could edit. The full study should validate against an immutable original test suite and check API compatibility, exceptions, performance, security, and untested inputs.
Finally, the manual validation was performed by the first author alone; the full study should use multiple annotators with agreement measures such as Cohen's kappa, and run the agent in a sandbox.

\paragraph{Lessons}
\begin{itemize}
\item \textbf{Lesson 1---Available context shapes dependency discovery and reported dependency scope.} The experiment with context reported substantially more dependency-transformation candidates, whereas the run without context relied more heavily on repository search. The audit found no evidence that the agent inspected the installed package environment, suggesting that its reported scope was primarily determined by the repository information made visible to it.

\item \textbf{Lesson 2---Dependency discovery may depend on the downstream goal.} We find that more inventory-oriented runs reported more dependency-transformation candidates than removal-oriented runs selected as transformation candidates. Agents may define dependency scope differently depending on whether they are asked to describe, remove, replace, or repair dependencies. Dependency analyses should therefore state the intended downstream purpose and evaluate goal sensitivity through matched and repeated runs.

\item \textbf{Lesson 3---Repair efficiency may require multiple cycles.} Dependency count, transformation-entry count, agent turns, and transformation operations represent different aspects of repair effort. A single transformation may resolve several transformation entries, so such transformations may be required to evaluate the cost.

\item \textbf{Lesson 4---Dependency transformation extends beyond local reimplementation.} The observed strategies vary and should be considered by developers and agent users.

\item \textbf{Lesson 5---Transformation completion does not independently establish behavioral preservation.} Although the audit found no deleted test files or newly introduced skips, many repair steps modified or added test artifacts, and at least one transformation changed an original behavioral expectation. In addition, in one attempt without context, a broader test run failed; the agent judged the failure to be unrelated to its change and completed the attempt after validating with a narrower set of tests.
\end{itemize}

\section{Research Agenda}

We now present a five-part research agenda for future work.

\paragraph{Contracts for Dependency Transformations}
\emph{The first part of the agenda investigates the types of contracts necessary to ensure validation.}
Empirical and simulation studies are needed to explore what contract details are needed to create measurable and comparative metrics between transformation activities.

\paragraph{Dependency Transformation Taxonomy}
\emph{The second part of the agenda develops a dependency transformation taxonomy beyond a single case study.}
We plan to develop a full empirical study that will present a comprehensive taxonomy can be consistent over programming language, repository types and AI models. 
This is immediate future work. 

\paragraph{Validation Beyond Build and Tests}
\emph{The third part of the agenda concerns validation of the transformation.}
Empirical studies will include exploring metrics and strategies to identify behavioral, environment-level, and accounting.
These studies will include human validation and surveys to understand the cost-effectiveness of these alternatives to dependency adoption and maintenance. 

\paragraph{Repair-Oriented Agent Workflows}
\emph{The fourth part of the agenda investigates different repair policies during dependency transformation.}
During dependency transformation, repair of the code is unavoidable. Therefore empirical studies and simulation studies are needed to evaluate effective methods to balance repair depth vs. cost. 
As agent pipelines become more common, strategies for cost-effective and reproducible studies are needed.

\paragraph{Human Checkpoints and Long-Term Trade-offs}
\emph{The fifth part of the agenda concerns human acceptance of dependency transformation and its long-term feasibility.}
These studies will need to be in the form of interviews and developer perspectives to understand whether a developer is willing to consider these transformations as an alternative to current dependency maintenance activities. 

\section{Conclusion}
These early observations motivate, rather than validate.
Even as a single case study, we show that using an agent opens up new alternatives for developers to consider in dependency management.

\section{Artifact Availability}
We make our replication artifact that contains prompts, scripts and execution logs for our experiment. 
\url{https://doi.org/10.5281/zenodo.23230583}

\section*{Acknowledgment}
This work was supported by JSPS KAKENHI Grant Number JP24H00692, JP25K03102, JP26K02889, JP26H02500, JP23K28065.

\bibliographystyle{IEEEtran}
\bibliography{bibliography}

\end{document}